# Additive manufacturing of Ni-Mn-Sn shape memory Heusler alloy – Microstructure and magnetic properties from powder to printed parts

*Franziska Scheibel*[1,*], *Christian Lauhoff*[2], *Philipp Krooß*[2], *Stefan Riegg*[1], *Niklas Sommer*[3], *David Koch*[4], *Konrad Opelt*[1,5], *Heiner Gutte*[6], *Olena Volkova*[6], *Stefan Böhm*[3], *Thomas Niendorf*[2], *Oliver Gutfleisch*[1]

[1] Functional Materials, Institute of Materials Science, Technical University of Darmstadt, Alarich-Weiss-Straße 16,
  64287 Darmstadt, Germany
[2] Institute of Materials Engineering, University of Kassel, Moenchebergstr. 3, 34125 Kassel, Germany
[3] Institute for Production Technologies and Logistics, University of Kassel, Kurt-Wolters-Str. 3, 34125 Kassel,
  Germany
[4] Structure research, Institute of Material Science, Technical University of Darmstadt, Alarich-Weiss-Straße 2,
  64287 Darmstadt, Germany
[5] Fraunhofer Institution for Material Cycles and Resource Strategies IWKS, Hanau, Germany
[6] Institute of Iron and Steel Technology, TU Bergakademie Freiberg, Leipziger Str. 34, 09599 Freiberg, Germany

*Corresponding author: franziska.scheibel@tu-darmstadt.de

**Abstract**:

Ni-Mn-based Heusler alloys like Ni-Mn-Sn show an elastocaloric as well as magnetocaloric effect during the magneto-structural phase transition, making this material interesting for solid-state cooling application. Material processing by additive manufacturing can overcome difficulties related to machinability of the alloys, caused by their intrinsic brittleness. Since the magnetic properties and transition temperature are highly sensitive to the chemical composition, it is essential to understand and monitoring these properties over the entire processing chain. In the present work the microstructural and magnetic properties from gas-atomized powder to post-processed Ni-Mn-Sn alloy are investigated. Direct energy deposition was used for processing, promoting the evolution of a polycrystalline microstructure being characterized by elongated grains along the building direction. A complete and sharp martensitic transformation can be achieved after applying a subsequent heat treatment at 1173 K for 24 h. The Mn-evaporation of 1.3 at. % and the formation of Mn-oxide during DED-processing lead to an increase of the transition temperature of 45 K and a decrease of magnetization, clearly pointing at the necessity of controlling the composition, oxygen partial pressure and magnetic properties over the entire processing chain.

# 1. Introduction

The functional properties of Ni-Mn-based Heusler alloys attract a lot of interest for a variety of different fields of research such as magnetic shape memory alloys [1–3], magnetocaloric [1,4–6], elastocaloric [7–9] and multicaloric materials [10–13] which can be potentially applied in actuators [14,15], for energy harvesting [16–18] or solid-state caloric cooling [19–22].

Ni-Mn-based Heusler alloys including Ni-Mn-Sn exhibits a first-order magneto-structural transformation (FOMST) from high-temperature ferromagnetic austenite to low-temperature paramagnetic martensite [6,23]. The FOMST can be induced by different stimuli, i.e. magnetic field, stress, hydrostatic pressure or temperature change [8,11,20,24]. The transition temperature can be adjusted via the chemical composition which makes the material an excellent candidate for "multi-stimuli caloric cooling" [20]. Unfortunately, Ni-Mn-Sn as well as other Ni-Mn-based Heusler alloys suffer extreme brittleness and, thus, are highly difficult to machine. To obtain near net shape parts of these alloys, recently, additive manufacturing came into focus of academia. Complex geometries of polycrystalline Ni-Mn-Ga and Ni-Mn-Sn alloys were successfully processed by laser powder bed fusion (PBF-LB/M) technique [25–27] or powder bed binder jetting [28,29]. Multicaloric applications, require fully dense structures to maximize the amount of active caloric material and, hence, to achieve a maximum in performance. From this, the additive manufacturing techniques to produce Ni-Mn-Sn based parts without binder material are PBF-LB/M or direct energy deposition (DED). Nonetheless, the set of processing parameters (e.g. laser power, layer thickness, hatching distance) to control the porosity and the composition still has to be investigated for highest density with optimal chemical composition [30]. The high vapor pressure of Mn makes it difficult to control the alloy composition, directly affecting the transition temperature and magnetic properties [25]. Thus, it is essential to monitor and investigate the Mn-content and transition temperature along the entire processing chain, i.e. from powder production to post-processed printed parts. Despite this challenge additive manufacturing can provide a further advantage: The possibility for a direct microstructure design. In Co-Ni-Ga shape memory Heusler alloys, a favorable bamboo like grain structure [31] or even highly anisotropic, textured microstructure, with functional properties close to single crystalline material [32] can be attained by PBF-LB/M and DED, respectively, leading to excellent mechanical functional properties. Indeed, a comparison between PBF-LB/M and DED-processed Co-Ni-Ga shows that the FOMST and magnetic properties of the DED-processed material are close to the behavior of single crystals [33].

A study on DED-processing of Ni-Co-Mn-Sn using ground non-spherical (irregular shaped) powder [34] reported that the functional properties of the alloy could be realized after subsequential heat treatment. The heat treatment for homogenization was required, since the as-processed alloy show an inhomogeneous microstructure, in addition a slight variation of the transition temperature was observed due to the Mn evaporation during processing. Since, the microstructure is sensitive to the processing parameters and scan-strategy a consistent powder flowrate is essential to achieve reproducible processing-conditions. Therefore, the use of spherical powder with homogeneous flowability is essential to study the functional and microstructural properties of additive manufactured Ni-Mn-Sn shape memory Heusler alloys. The powder production is a challenge since rather large quantities of powder are required for PBF-LB/M (around 10-20 kg) just for the process qualification. The alloys are not commercially available and the usual lab scale for conventional processing (arc-melting [6,35] or induction melting [5,36]) is in the range of several gram [37].

In the present study, spherical powder produced by vacuum inert gas-atomization (VIGA) was used for DED-processing of a Ni-Mn-Sn Heusler alloy. A detailed powder characterization regarding size, shape, chemical composition, and homogeneity is shown and the functional properties are reported for both powder and additively manufactured Ni-Mn-Sn in as-processed as well as post-processed state, respectively. The FOMST and magnetic properties of DED-processed Ni-Mn-Sn alloy are also compared to conventional processed bulk material. The study covers the entire process chain from raw material over spherical powder production towards additive manufacturing and post-processing of Ni-Mn-Sn shape memory Heusler alloys including the monitoring of structural, microstructural, compositional, and magnetic properties for each different processing step. By this it is possible to identify the important process parameters along the processing chain defining the functionality of the final Ni-Mn-Sn alloy.

## 2. Materials and Methods

The raw material for powder production was supplied by *Less Common Metals Ltd.*: $Ni_{50}Mn_{50}$ pre-alloy and Ni as well as Sn pellets with a purity of 99.9%, 99.99% and 99.9%, respectively. For fabrication of spherical shaped $Ni_{49.8}Mn_{36.6}Sn_{13.6}$ powder, 0.98 kg Ni, 4.30 kg $Ni_{50}Mn_{50}$ and 1.72 kg Sn were used for VIGA at TU Bergakademie Freiberg. For additive manufacturing, powder material featuring a particle size between 20 µm and 50 µm was obtained by sieving using a compact screening machine *KSM 500* from *Assonic* with ultrasound stimulation. Screen clothes with a mesh size of 20 µm and 50 µm were used. To minimize the content of fine particles the powder was sieved twice. The particle size distributions were determined by laser diffraction using a *Mastersizer 3000* from *Malvern Instruments Ltd.*. Based on the rotating drum principle the cohesion and first avalanche angle within the powder was analyzed using a *GranuDrum* from *Granutools*. The results are presented in the supplementary information.

The sieved powder was processed by DED technique on a 773 K preheated Ni-Mn-Sn substrate material. The powder was fed by a two-channel powder feeder to a three-jet-nozzle; Argon gas was used as carrier gas as well as shielding gas to prevent oxidation. The powder was melted by a 2 kW multi-mode fiber-laser with a wavelength of 1070 nm (*YLS-2000-S2, IPG Photonics*) employing a quasi-bidirectional scanning strategy to process cuboidal blocks. The nominal dimensions of the cuboids and the processing parameters are given in Tab. 1. Fig. 1 shown an image illustrating a processed cuboid, the underlying Ni-Mn-Sn substrate, and the steel building platform. The temperature of the building platform and the Ni-Mn-Mn substrate was kept at 773 K during processing. The Ni-Mn-Sn substrate material was used to avoid cross-contamination between the steel building platform and the processed cuboid. The temperature was determined by a thermocouple placed between building platform and Ni-Mn-Sn substrate. Fig. 1 also shows a schematic of the scanning strategy. In the remainder of the text, the building direction (BD) is the direction perpendicular to the building platform and the scanning direction (SD) is the direction of the first layer, while the SD is rotated by 180° for each layer (quasi-bidirectional scanning strategy). The 10 x 10 x 8 mm cuboids are built from 10 tracks with a track spacing of 1 mm and 27 layers with a layer distance of 0.3 mm.

For more details on the DED setup and the processing parameters used the reader is referred to [32,38]. The Ni-Mn-Sn substrate was prepared by arc melting, i.e., 80 g of $Ni_{50}Mn_{36}Sn_{14}$ using the same raw materials as for VIGA powder was molten in ingot form. The top and bottom surfaces were cut parallel by electro-discharge machining (EDM). Three different laser powers were used to study their influence on the evaporation of Mn. Inductively coupled plasma optical emission spectrometry (ICP-OES) was performed for compositional analysis of the three samples processed with laser powers of 400 W, 500 W, and 600 W to determine the Mn loss due to evaporation during processing (Tab. 3). The spectrometry was performed using an *iCAP PRO* ICP-OES system from *ThermoFischer*.

*Tab. 1: Summary of the DED-processing parameters used in present work and the chemical composition of the samples after processing determined by ICP-OES.*

| Dimensions (L x W x H) | 10 x 10 x 8 mm³ |
|---|---|
| Mass flow | 4 g/min |
| Track spacing | 1 mm |
| Number of tracks | 10 |
| Layer distance | 0.3 mm |
| Number of layers | 27 |
| Feed rate | 10 mm/s |
| Platform temperature | 773 K |
| Laser power | 400 W, 500 W, 600 W |

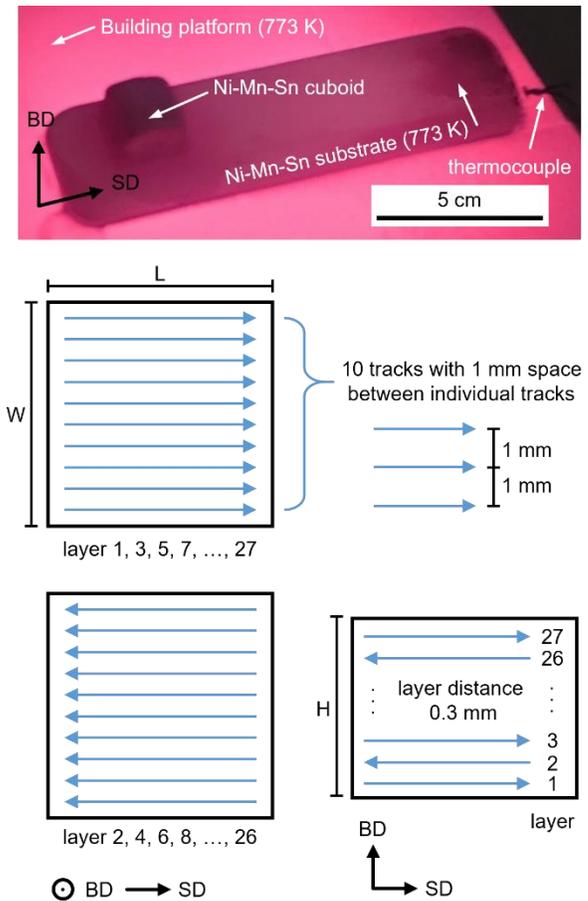

*Fig. 1: Image showing a DED processed Ni-Mn-Sn cuboid. The processing was performed on a Ni-Mn-Sn substrate at 773 K. BD and SD of the first layer are marked by arrows. SD is turned by 180° after each layer. A sketch of the scan strategy is also shown below, parallel (left) and perpendicular (right) to BD.*

For the subsequent heat treatment of the DED-processed bulk alloy and the gas-atomized powder, the bulk sample and powder were encapsulated in quartz-glass under Argon atmosphere and annealed at 1173 K for 24 h and 6 h, respectively. After annealing, the quartz capsules were quenched in water. Local chemical composition and homogeneity of the powder and the DED-processed bulk material before and after the heat treatment was determined by energy-dispersive x-ray spectroscopy (EDS) using a *Tescan Vega 3* scanning electron microscope (SEM) with an *EDAX Octane Plus* detector. The composition was also determined with ICP-OES for a more accurate but global composition analysis. For microstructure analysis scanning electron microscopy (SEM) including electron backscatter diffraction (EBSD) was employed using a *TESCAN CamScan MV 2300* and a *Zeiss Ultra Plus Gemini* microscope. All measurements were performed at room temperature.

Temperature dependent x-ray diffraction (XRD) measurements were performed using Mo Kα radiation in transmission geometry on a custom-built setup with a *Mythen2 R 1K* detector. The powder was mixed with *NIST 640d* standard silicon powder for correcting geometric errors and glued on a graphite foil. The measurements were carried out during cooling. The data analysis was done with *JANA2006* [39] within the superspace approach [40].

Temperature and field-dependent magnetization measurements in the temperature range between 10 K and 380 K were performed utilizing a Magnetic Property Measurement System (MPMS) from *Quantum Design* equipped with a 7 T field and a vibrating sample magnetometer (VSM). The critical temperatures of the FOMST ($A_s$, $A_f$, $M_s$, and $M_f$) were determined by the tangent method [6]. The thermal hysteresis $\Delta T_{hyst}$ and the Curie temperature $T_C$ of the austenite phase were determined by the maximum and minimum of the derivative $dM(T)/dT$, respectively.

## 3. Results and Discussion
### 3.1. Powder characterization
#### 3.1.1. Microstructure and composition

Fig. 2 shows the SEM backscattered election (BSE) images of the as-atomized and the sieved powder. In case of the as-atomized powder spherical particles, partially with irregular shape can be detected. The spherical particles exhibit satellites on the particle surface, marked by the red arrows in Fig. 2a). As well-known from literature, such satellite particles are formed by collisions during solidification [41]. Larger particles (> 50 µm) as well as small particle sizes below 20 µm are separated by sieving. In the sieved powder, shown in Fig. 2b), a residual number of particles below 10 µm remains, which is also confirmed by the particle size distribution shown in Fig. 2c). The size distribution of both powders shows a median ($D_{50}$) of 33 µm. The as-atomized powder has a wide Gaussian-type particle size distribution, i.e., 90% of the powder particles have a size below 92 µm ($D_{90}$) while 10% are below 10 µm ($D_{10}$). In turn, sieving leads to values of $D_{90}$ = 56 µm and $D_{10}$ = 17 µm, i.e., a narrow size distribution. However, a second peak in the distribution is visible at approximately 10 µm. The very fine particles can agglomerate or attach to the larger particles due to high cohesiveness, explaining their appearance after sieving.

Fig. 3a) shows a cross-section SEM-BSE image of the particles with a size between 20 µm and 50 µm after the VIGA process. The dendritic microstructure hints at a chemically inhomogeneous character of the particles [42]. Thus, a subsequent heat treatment is required to compare their magnetic and structural properties with values reported in literature. A heat treatment of 6 h at 1173 K is sufficient to homogenize the particles. Note, the relatively short annealing time compared to the bulk material (24 h at 1173 K, cf. Section 2) has been chosen to prevent the powder particles from sintering. The cross-section SEM-BSE image of the heat-treated powder in Fig. 3b) shows homogeneous polycrystalline particles. EDS analysis of the particle cross-sections reveal the same chemical composition for the individual particles before and after heat treatment. The average composition of the particle cross-section and the related standard deviation of the 20 µm to 50 µm size particles is shown in Tab. 2. The low standard deviation proofs that the individual particles are characterized by negligible variations in the chemical composition. Since the error of EDS measurements can be usually up to 2 at. % with respect to absolute values, the stoichiometry is also determined by ICP-OES analysis (Tab. 2). The comparison between as-atomized, sieved, and sieved and heat-treated powder reveals no variations in the composition.

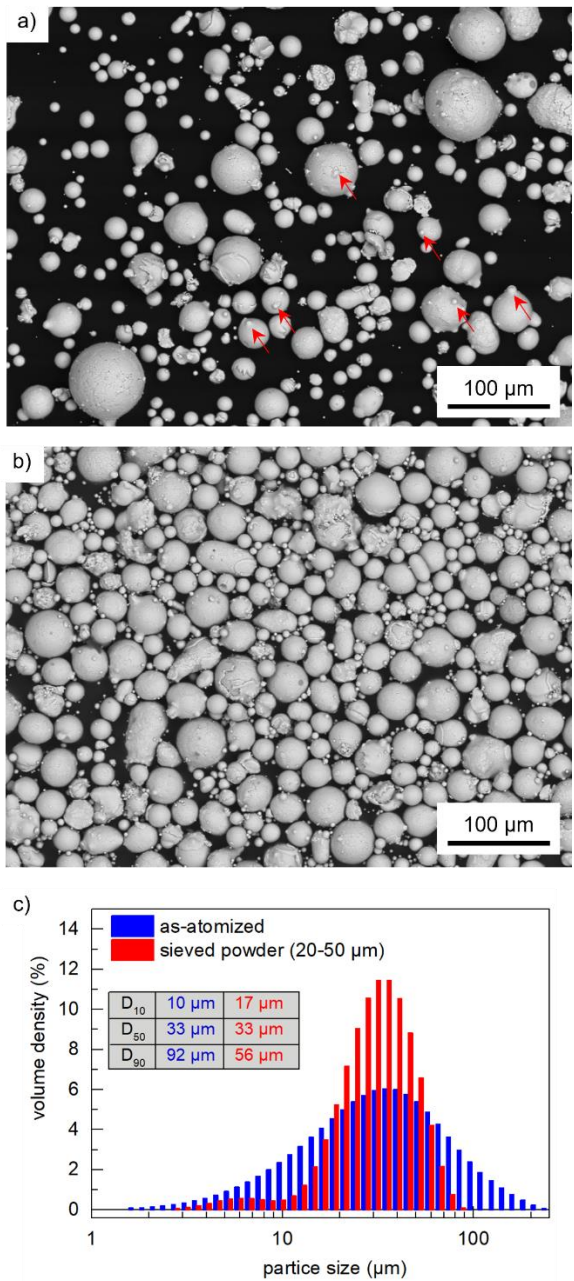

Fig. 2: BSE images of the (a) as-atomized and (b) sieved powder with a particle size between 20 µm and 50 µm. (c) Particle size distribution of as-atomized and sieved power including median $D_{50}$, $D_{10}$ and $D_{90}$ values measured by laser diffraction.

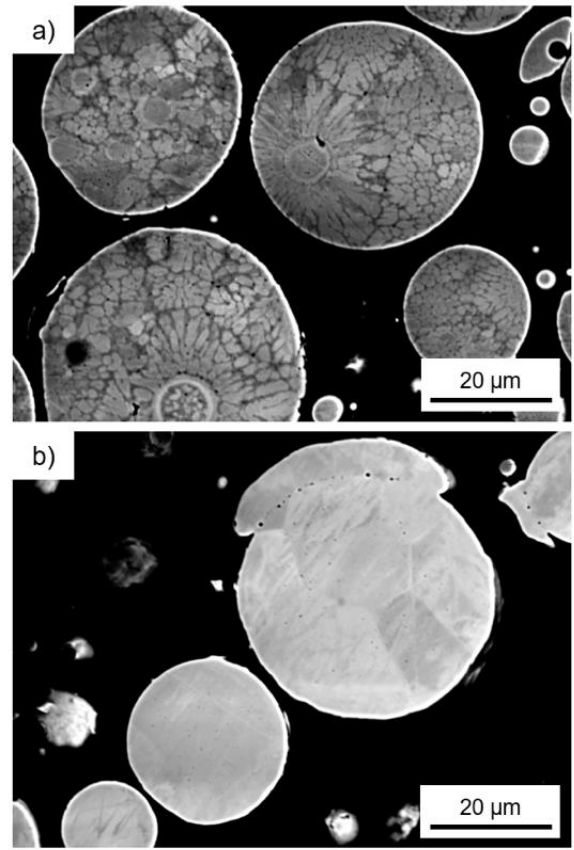

Fig. 3: Cross-section BSE images of the powder size between 20 µm and 50 µm a) before and b) after heat treatment at 1173 K for 6 h.

*Tab. 2: Chemical composition of gas-atomized powder before and after sieving and after subsequent heat treatment studied by ICP-OES and EDS. EDS are performed on the particles cross-section, the average values of the individual particle measurements and the corresponding standard deviation (Std.) are shown.*

| Powder | ICP-OES (at. %) | | | | EDS on particles cross-sections (at. %) | | | |
|---|---|---|---|---|---|---|---|---|
| | Ni | Mn | Sn | e/a | Ni | Mn | Sn | Std. |
| as-atomized | 49.4 | 36.0 | 14.6 | 8.04 | 48.7 | 37.8 | 13.5 | 0.3 |
| 20 µm to 50 µm | 49.4 | 36.1 | 14.5 | 8.05 | 48.9 | 37.6 | 13.5 | 0.4 |
| 20 µm to 50 µm 1173 K 6 h | 49.5 | 36.0 | 14.5 | 8.05 | 48.8 | 37.6 | 13.6 | 0.4 |

### 3.1.2. First-order magneto structural transformation

The phase composition of the particles with a size between 20 µm and 50 µm is investigated in the as-atomized and heat-treated condition using x-ray diffraction in a temperature range from 25 K to 450 K. An overview about of all measured temperatures can be found in the appendix (S2). For both powders, as-atomized and heat treated, no evidence of Mn-oxide phase is detected, proving the high quality of the powder. Fig. 4a) and c) depicts exemplarily the Rietveld refinement in the as-atomized condition, performed at 25 K and 450 K, respectively. At 25 K, 4O and $L1_0$ martensite as well as residual austenite are present. The temperature dependent phase fraction analysis in Fig. 4b) shows that the martensitic phase (4O and $L1_0$) gradually transforms into the austenitic phase with increasing temperature (cyan area). However, a small amount of residual martensite is present even at 450 K, this is recognizable in Fig.4c) at the shoulder of the (220) peak at ~19°. The large temperature span of the transformation of about 200 K as well as the residual martensite and austenite at high and low temperatures, respectively, are in good agreement with the inhomogeneous microstructure found for the particles (cf. Fig. 3a)). In contrast, the diffraction pattern at 25 K of the heat-treated particles shown in Fig. 4d) indicate no presence of $L1_0$ martensite phase and no residual austenite. The phase fraction analysis (Fig. 4e) exhibits no residual austenite phase below 100 K and only a small amount (<10%) of austenite phase up to 240 K. Between $M_s \approx$ 240 K and $M_f \approx$ 280 K a transformation from martensite to austenite can be observed (cyan area). Above 280 K no residual martensite can be observed. Fig. 4f) shows the Rietveld refinement at 400 K, indicating no presence of residual martensite phase. The sharp FOMST in the heat-treated powder indicates a simultaneous transformation behavior of the particles, occurring when the composition of the individual particles is the same (Tab. 2).

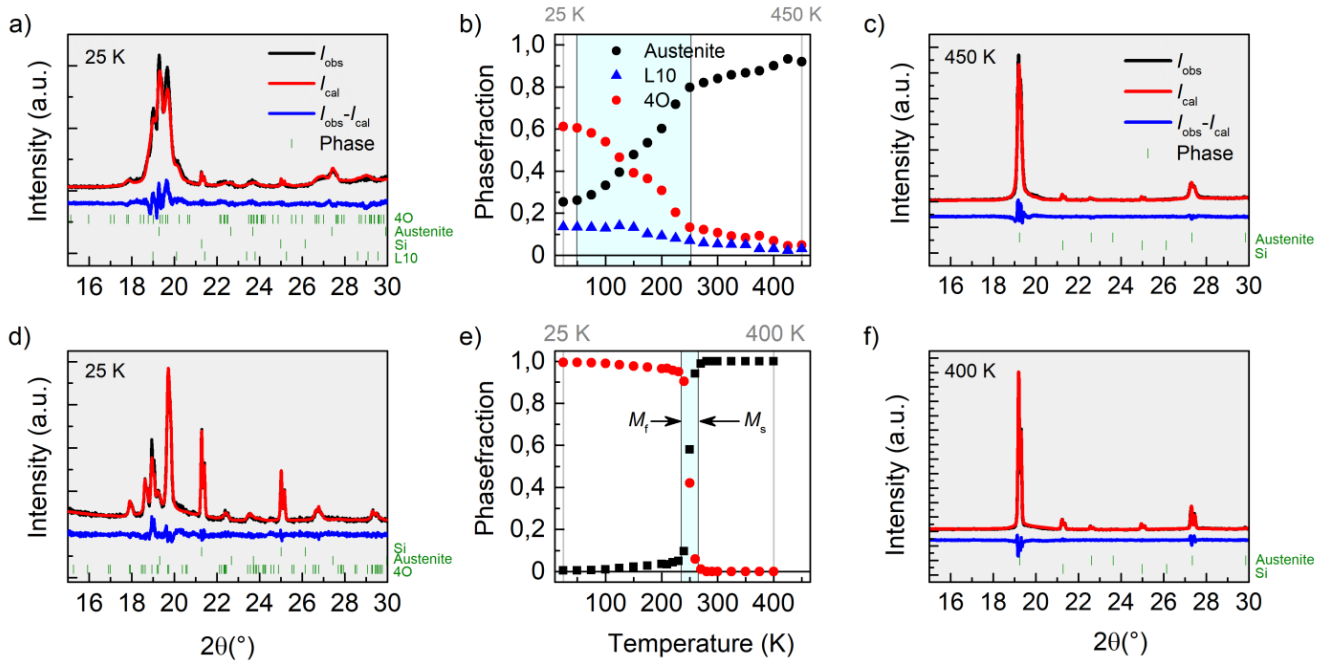

*Fig. 4: X-ray diffraction of Ni-Mn-Sn gas-atomized powder with a particle size between 20 µm and 50 µm. Rietveld refinement obtained at 25 K and 450 K for a) and c) the as-atomized and d) and f) heat-treated condition, respectively. The recorded diffraction pattern, the refinement and their deviation are shown in black, red, and blue, respectively. For the refinement, 4O and $L1_0$ martensite, austenite as well as Si as reference standard are used. b) and e) show the temperature-dependent phase fraction evolution of b) the as-atomized and e) heat-treated condition.*

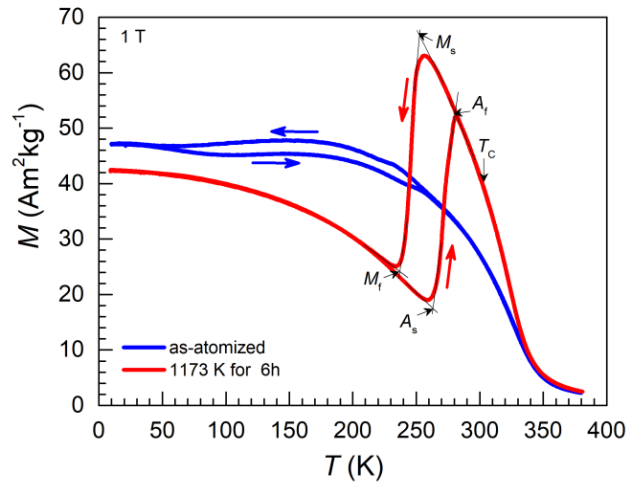

Fig. 5: Temperature-dependent magnetization of sieved Ni-Mn-Sn powder (20 µm to 50 µm) with and without heat treatment 1173 K for 6 h. Measurements are performed in a constant field of 1 T using field cooling and heating protocol.

The sharp and complete FOMST of the homogenized powder is also confirmed by magnetic characterization. Fig. 5 shows the temperature-dependent magnetization for the powder before and after heat treatment, measured in 1 T. The splitting between the field cooling and field heating $M(T)$ curve (measuring direction is indicated by arrows) is related to the first-order nature of the transformation. Before heat treatment, the inhomogeneous powder exhibits a transformation over a large temperature range between 50 K and 250 K, which is consistent with the transformation behavior determined by x-ray diffraction phase analysis in Fig. 4b). In comparison, the heat-treated powder is characterized by a sharp FOMST ($M_f$ = 237 K, $M_s$ = 252 K, $A_f$ = 281 K, and $A_s$ = 263 K) with a large change in the magnetization $\Delta M$ = 37.9 Am$^2$kg$^{-1}$ and a thermal hysteresis of $\Delta T_{hyst}$ = 23.2 K. The Curie temperature of the austenite phase is determined at $T_C$ = 325 K. The magnetic behavior of the homogenized powder is comparable with bulk material featuring similar composition [1,6,43,44]. This clearly proves that the functional properties of inhomogeneous gas-atomized powder can be regained by post-process heat treatment. Thus, the precursor material for the DED-processing of Ni-Mn-Sn shows no deterioration in magnetic properties compared to bulk samples from the literature.

## 3.2. Additive manufacturing by direct energy deposition

For DED-processing, the sieved powder with particle size between 20 µm and 50µm is used in the as-atomized condition. The inhomogeneity of the individual particles is not an issue since the used laser powers are high enough to melt the entire particle. The chemical composition of the specimens processed with 600 W, 500 W and 400 W is determined by ICP-OES, the results are shown in Tab. 3. The sample processed with 600 W exhibits a significant lower Mn content than the samples processed with lower laser power (500 W and 400 W). This general trend is also observed in Ni-Mn-Ga Heusler alloys [27,45], and is related to the increased Mn evaporation with increasing power. The sample processed with the lowest laser power (400 W) exhibits the lowest porosity and a chemical composition closest to the one of the heat-treated powder. Therefore, this sample is selected for further investigations of the microstructure and magnetic properties.

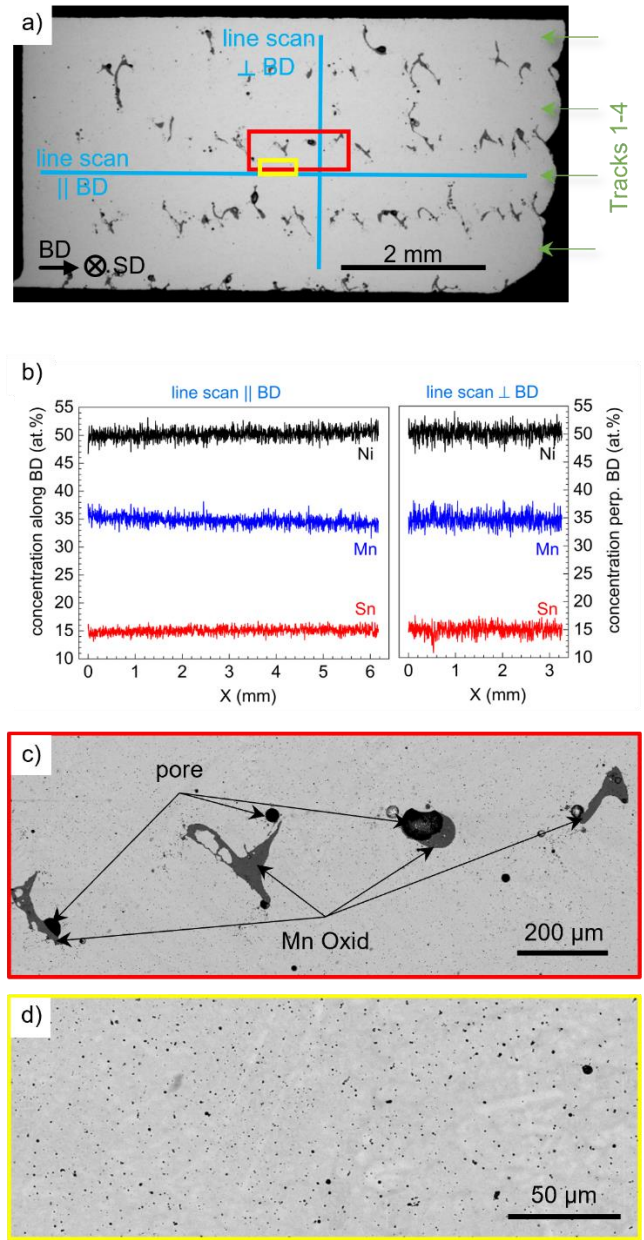

*Tab. 3: Chemical composition and e/a ratio of the samples after DED-processing determined by ICP-OES.*

| Laser power | ICP-OES (at. %) | | | e/a |
|---|---|---|---|---|
| | Ni | Mn | Sn | |
| 600 W | 52.9 | 31.8 | 13.3 | 8.13 |
| 500 W | 50.4 | 34.8 | 14.8 | 8.07 |
| 400 W | 50.5 | 34.8 | 14.7 | 8.07 |

### 3.2.1. Microstructural and structural analysis

The microstructural of the as-processed Ni-Mn-Sn material is investigated using SEM including BSE and EDS line scans. Fig. 6 shows the analysis of a cross-section parallel to BD and perpendicular to SD. The BSE images in Fig. 6a), shows that the 400 W-processed material features some amount of porosity and secondary phase after processing. Future work is already in progress to fabricate full dense material being free of cracks and pores. However, it should be emphasized that the pores and cracks shown in Fig. 6 have no influence on the specific magnetization (Sec. 3.2.2), since the magnetic moment is related to the mass.

*Fig. 6: a) BSE image and b) EDS line scans of DED-processed Ni-Mn-Sn cross-section parallel to BD. c) and d) show the BSE images with higher magnification of the red and yellow marked areas in a), respectively.*

EDS line scans are performed to investigate the chemical composition parallel and perpendicular to BD, the line scans are marked in Fig. 6a) by the vertical and horizontal blue lines. The EDS line scan perpendicular to BD is shown in Fig. 6b) on the right sight, indicating no variation of the composition. The line scan crossed four tracks marked by green arrows indicating a homogeneous composition across different tracks. The line scan parallel to BD, shown at the left sight in Fig. 6b), reveal a slight decrease of the Mn content with increasing layer number. However, the variation is very small and lies within the scatter of the measured values. A possible cause for the decreasing Mn content could be an increasing Mn evaporation with increasing distance to the heated Ni-Mn-Sn substrate. since the temperature of each layer could not be determined during processing a temperature gradient parallel to the building direction cannot be excluded.

Fig. 6a) shows large defects in the area where two tracks are overlapping. The magnification of such an area marked by the red rectangle is shown in Fig. 6c), revealing large pores and a Mn-oxide phase. The presence of Mn-oxide is confirmed by EDS analysis. An EDS mapping is also shown in the supplementary (S 3). Mn-oxide phases can affect the magnetic properties, and decrease the performance of magnetocaloric materials [46]. In contrast, the center of the track is free of this up to 200 µm large defects. The magnification of such an area marked by the yellow rectangle is shown in Fig. 6d). In the center of the track only small pores (black contrast) with a size of approximately 1 µm can be observed. The origin of the defects are diverse, the larger defects at the edge of the track are most probably related to a lack of fusion, while the smaller pores at the center of the track are most probably caused by entrapped gas [47]. In Fig. 6d), the brighter contrast indicates an inhomogeneous chemical composition. EDS point analysis of the light gray areas confirm a Sn-rich Ni-Mn-Sn phase while the dark gray areas are Sn-poor. The sample is homogenized by subsequent heat treatment at 1173 K for 24 h (see supplementary S2). The treatment successfully removed the local chemical inhomogeneities (Fig. 6d) as well as the chemical gradient parallel to BD (Fig. 6b). Even smaller defects of Mn-oxide phase up to 1 µm can be healed by a heat treatment of 24 h [26]. However, the heat treatment is not able to eliminate the large defects shown in Fig. 6c) (s. supplementary data S 2c)). A modification in the scan strategy using a 90° rotation of SD between alternating layers can potentially eliminate the larger defects in the overlap region of two adjacent tracks. A further possibility to reduce defects and pores in additive manufactured specimens is subsequent hot isostatic pressing (HIP) [48]. Since no Mn-oxide phases are detected in the XRD measurements (Fig. 4), the oxygen input into the specimen is almost certainly due to the residual oxygen content in the DED-chamber and leads to the formation of Mn-oxides during processing. A reduction of the residual oxygen content can therefore reduce the amount of Mn-oxide phase in the DED-processed material.

The grain size and orientation of the 400 W-processed Ni-Mn-Sn material are investigated by EBSD analysis at room temperature. Fig. 7 a) and c) show the color-coded inverse pole figure (IPF) maps of the cross-section parallel to BD for the as-built and heat-treated state, respectively, color coding is indicated by the standard triangle in Fig. 7d). The indexed grains correspond to the austenite phase with cubic crystal structure (*Fm-3m*) and a lattice parameter of 0.5986 nm. Fig. 7a) reveals a polycrystalline microstructure with elongated grains along BD. The grains have a length of 50 µm to 200 µm and a width of 20 to 50 µm. The black areas result from pores and phases, which do not relate to the austenite phase. The non-indexed phases are mainly located at the grain boundaries but are also found in the grain. The EBSD image quality (IQ) map in Fig. 7b) shows a lower quality of the Kikuchi-lines mainly at the grain boundaries. The IQ contrast results from the inhomogeneous nature of the as-processed Ni-Mn-Sn, shown also in Fig. 6d). Since the FOMST increases with decreasing Sn-content [6,43], regions with less Sn are at room temperature in the martensite phase. Similar EBSD analysis are also found in dual phase ferrite and martensite steels [49]. The subsequent post-process heat treatment causes a homogenization of the material. In consequence, the IPF map in Fig. 7c) reveals no traces of martensite phase or other non-indexed phases, except for the spherical black areas caused by pores. Besides the homogenization of the matrix, no considerable grain growth can be observed after the heat treatment at 1173 K for 24 h. Regarding the grain orientation no preferred orientation along BD can be observed.

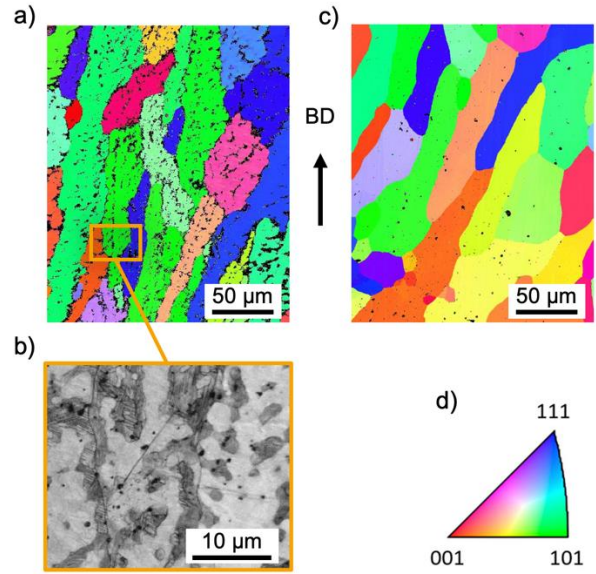

*Fig. 7: EBSD analysis of DED processed Ni-Mn-Sn: IPF mappings plotted with respect to BD a) before and b) after post-process heat treatment. BD is vertical and the color-coded standard triangle is shown in d). b) IQ image of the marked area in a) showing the inhomogeneous microstructure with Sn-rich (bright) and Sn-poor (dark) phases.*

### 3.2.2. Martensitic transformation behavior of the build and heat-treated bulk material

The temperature-dependent magnetization curves of the as-built and heat-treated Ni-Mn-Sn sample processed with 400 W laser power are shown in Fig. 8. For better comparison, the $M(T)$-curve of the annealed powder shown in Fig. 5 (light red) and a Ni-Mn-Sn bulk sample (dashed gray) are shown as well. The magnetization measurements of the $Ni_{51.2}Mn_{35.1}Sn_{13.7}$ bulk sample are adapted with permission of A. Taubel form Ref. [6], this bulk sample was selected because its transition temperature is comparable to that of the heat-treated DED-processed sample. The DED-processed sample in the inhomogeneous as-built condition is characterized by a flat curve (blue curve in Fig. 8). The broadening of the transition is also known for other Heusler alloys in the as-build or as-cast state [6,32]. The slight splitting of the heating and cooling curves indicates a partial magneto-structural transformation. However, a sharp FOMST can be realized by the post-process heat treatment at 1173 K for 24 h (red). The transition temperature ($A_s$ = 285 K, $A_f$ = 298 K, $M_s$ = 284 K, and $M_f$ = 272 K) is closer to the Curie temperature $T_C$ = 301 K and exhibit a $\Delta T_{hyst}$ of 15.1 K and a $\Delta M$ of 11.7 $Am^2kg^{-1}$. The martensite transition temperature correlates with the chemical composition and increases with increasing e/a ratio, while $T_C$ of the austenite phase decrease with increasing e/a ratio [1,6,43,50]. The reduction of Mn content increases the e/a ratio and a leads to a destabilization of the ferromagnetic austenite phase with reduced saturation magnetization since the magnetic moment is carried mainly by the Mn-atoms [51]. Due to the Mn losses during DED-processing, the additively manufactured Ni-Mn-Sn alloy exhibit a higher e/a ratio compared to the Ni-Mn-Sn powder (Tab. 2 and 3), which leads to a 46 K higher FOMST temperature and a 24 K lower Curie temperature compared to the powder material. The comparison between heat-treated DED-processed and bulk sample from Ref. [6] reveal a broader transition width and a lower magnetization of the austenite phase in case of the DED-processed sample, leading to a reduction of $\Delta M$. The magnetization of the austenite phase is related to the degree of atomic ordering [52], however, the presence of Mn-oxide phase in the DED-processed sample also effects the magnetization since the specific magnetization per mass includes the paramagnetic Mn-oxide phase. The width and thermal hysteresis of FOMST are also affected by defects, secondary phases, grain-size and internal stress [33,53–56]. The Mn-oxide phase as well as the pores also effect the transition process because the defects can act as pinning centers causing a broadening of the FOMST. In contrast the defect free bulk sample exhibit a sharp FOMST. The reduction of pores and Mn-oxide is therefore a goal to achieve the same magnetic properties and high magnetocaloric effect as in conventional bulk material. To compensate the Mn losses during processing, a Mn excess can be added for powder production. A further possibility to increase the magnetization is the addition of small amounts of Co (~5 at. %) [6,57]. It is shown that this increases the magnetization and stabilizes the austenite phase. In addition, the Curie temperature is shifted towards higher temperatures [58], which allows for the compensation of the observed shift due to the DED-processing.

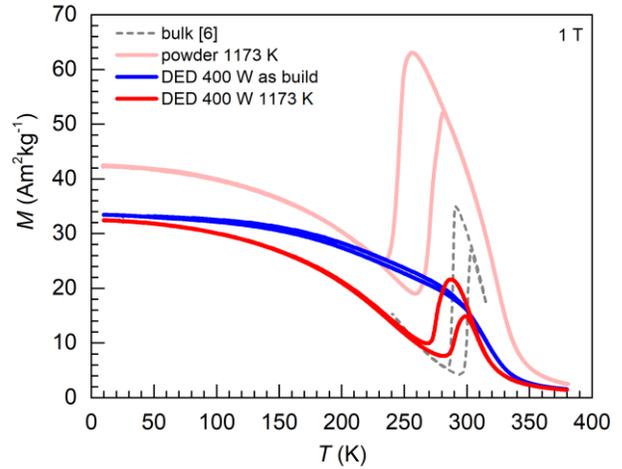

*Fig. 8: Temperature-dependent magnetization of DED processed Ni-Mn-Sn in the as-build and heat-treated condition. The measurements are performed in 1 T using field cooling and heating protocol. For comparison the M(T)-curve of the heat-treated powder (Fig. 5) and a Ni-Mn-Sn bulk sample, adapted with permission from A. Taubel Ref. [6], is shown as well.*

### 4. Summary and conclusions

In this study, the entire processing chain from powder preparation towards DED-processed Ni-Mn-Sn bulk material is reported, including post-processing, which turned out to be essential for achieving a complete and sharp martensitic transformation in the printed parts. The results of the detailed microstructural, crystal structural and phase analysis, compositional, and magnetic analysis of the gas-atomized powder and DED-processed material can be summarized as follows:

- After vacuum inert gas-atomization, powder with spherical particle shape, but inhomogeneous composition is obtained. The latter leads in the as-atomized state to a broad FOMST with residual martensite and austenite phases at high and low temperatures, respectively. The powder shows not indication of Mn-oxide phase.

- A post-process heat treatment at 1173 K for 6 h is sufficient to homogenise the powder and achieve a sharp FOMST that approximates the properties of conventionally cast bulk material [6]. The heat-treated powder undergoes a complete FOMST, no residual martensite and austenite phase are determined at temperatures above and below the FOMST, respectively.
- Additive manufacturing of the gas-atomized powder using DED technique leads to bulk material with lower Mn-contend compared to the powder precursor material due to evaporation of Mn, 4.3 at. % for 600 W and 1.3 at. % for 500 W and 400 W (Tab. 1). A similar amount of Mn-evaporation is also reported for Ni-Mn-Ga using PBF-LB/M technique [27].
- Post-processed DED-processed Ni-Mn-Sn (117 K for 24 h) exhibit a polycrystalline microstructure with elongated grains along BD, the sample show no texture. After post-processing the sample is homogenous and shows no variation of the chemical composition perpendicular or parallel to BD.
- The homogenization via subsequent heat treatment regains the sharp FOMST, however the transition temperature increases by 46 K compared to the initial powder, due to the Mn-evaporation.
- The comparison between DED-pressed Ni-Mn-Sn and bulk material reveal a reduction of $\Delta M$ and an increase of the width of the FOMST, due to the presence of defects like Mn-oxide secondary phase and pores formed during DED-processing.

By optimizing the process parameters defects and Mn-oxide phase can be reduced to achieve the same magnetic performance as conventionally casted Ni-Mn-Sn bulk material. In addition, doping of Ni-Mn-Sn with Co can be used to in increase the Curie temperature and increase $\Delta M$ at the FOMST around room temperature [6,58]. The sharp FOMST after heat-treatment and the microstructure with elongated grains along BD obtain the potential for DED-processed Ni-Mn-Sn for the manufacturing of multicaloric Heusler alloys. Further research is particularly concerned with the implementation of textures to achieve excellent mechanical integrity and cyclability, as has already been achieved in DED-processed Co-Ni-Ga Heusler alloys [32]. This, together with defect reduction, increases the multicaloric performance of Ni-Mn-Sn alloys, making this material a potential candidate for solid-state caloric cooling.

**Acknowledgements:**

This work is supported by funding from the European Research Council (ERC) under the European Union's Horizon 2020 research and innovation program (grant number. 743116 - Cool Innov), the Deutsche Forschungsgemeinschaft CRC/TRR 270 "HoMMage" Project-ID 405553726-TRR 270. TN acknowledges funding by DFG (grant number 456078747). FS thanks Dr. Tino Gottschall and Dr. Andreas Taubel for providing the data of the Ni-Mn-Sn bulk sample (Fig. 8) and Lukas Schäfer the help regarding the cohesion and first avalanche angle analysis.

**Data availability statement:**

The data that support the findings of this study are available from the corresponding author upon reasonable request.

**Declaration of Competing Interest:**

The authors declare that they have no known competing financial interests or personal relationships that could have appeared to influence the work reported in this paper.

**References:**

[1] A. Planes, L. Manosa, M. Acet, Magnetocaloric effect and its relation to shape-memory properties in ferromagnetic Heusler alloys, J. Phys. Condens Matter. 21 (2009) 233201. https://doi.org/10.1088/0953-8984/21/23/233201.

[2] K. Ullakko, J.K. Huang, C. Kantner, R.C. O'Handley, V.V. Kokorin, Large magnetic-field-induced strains in $Ni_2MnGa$ single crystals, Appl. Phys. Lett. 69 (1996) 1966–1968. https://doi.org/10.1063/1.117637.

[3] E. Faran, D. Shilo, Ferromagnetic Shape Memory Alloys—Challenges, Applications, and Experimental Characterization, Exp. Tech. 40 (2015) 1005–1031. https://doi.org/10.1007/s40799-016-0098-5.

[4] A. Planes, L. Mañosa, X. Moya, T. Krenke, M. Acet, E.F. Wassermann, Magnetocaloric effect in Heusler shape-memory alloys, J. Magn. Magn. Mater. 310 (2007) 2767–2769. https://doi.org/10.1016/j.jmmm.2006.10.1041.

[5] J. Liu, T. Gottschall, K.P. Skokov, J.D. Moore, O. Gutfleisch, Giant magnetocaloric effect driven by structural transitions, Nat. Mater. 11 (2012) 620–626. https://doi.org/10.1038/nmat3334.

[6] A. Taubel, T. Gottschall, M. Fries, S. Riegg, C. Soon, K.P. Skokov, O. Gutfleisch, A Comparative Study on the Magnetocaloric Properties of Ni-Mn-X(-Co) Heusler Alloys, Phys. Status Solidi B. 255 (2018) 1700331. https://doi.org/10.1002/pssb.201700331.

[7] Y. Shen, W. Sun, Z.Y. Wei, Q. Shen, Y.F. Zhang, J. Liu, Orientation dependent elastocaloric effect in directionally solidified Ni-Mn-Sn alloys, Scr. Mater. 163 (2019) 14–18. https://doi.org/10.1016/j.scriptamat.2018.12.026.

[8] W. Sun, J. Liu, B. Lu, Y. Li, A. Yan, Large elastocaloric effect at small transformation strain in $Ni_{45}Mn_{44}Sn_{11}$ metamagnetic shape memory alloys, Scr. Mater. 114 (2016) 1–4. https://doi.org/10.1016/j.scriptamat.2015.11.021.

[9] B. Lu, P. Zhang, Y. Xu, W. Sun, J. Liu, Elastocaloric effect in $Ni_{45}Mn_{36.4}In_{13.6}Co_5$ metamagnetic shape memory alloys under mechanical cycling, Mater. Lett. 148 (2015) 110–113. https://doi.org/10.1016/j.matlet.2015.02.076.

[10] L. Pfeuffer, J. Lemke, N. Shayanfar, S. Riegg, D. Koch, A. Taubel, F. Scheibel, N.A. Kani, E. Adabifiroozjaei, L. Molina-Luna, K.P. Skokov, O. Gutfleisch, Microstructure engineering of metamagnetic Ni-Mn-based Heusler compounds by Fe-doping: A roadmap towards excellent cyclic stability combined with large elastocaloric and magnetocaloric effects, Acta Mater. 221 (2021) 117390. https://doi.org/10.1016/j.actamat.2021.117390.

[11] A. Gràcia-Condal, T. Gottschall, L. Pfeuffer, O. Gutfleisch, A. Planes, L. Mañosa, Multicaloric effects in metamagnetic Heusler Ni-Mn-In under uniaxial stress and magnetic field, Appl. Phys. Rev. 7 (2020) 041406. https://doi.org/10.1063/5.0020755.

[12] Y. Qu, A. Gràcia-Condal, L. Mañosa, A. Planes, D. Cong, Z. Nie, Y. Ren, Y. Wang, Outstanding caloric performances for energy-efficient multicaloric cooling in a Ni-Mn-based multifunctional alloy, Acta Mater. 177 (2019) 46–55. https://doi.org/10.1016/j.actamat.2019.07.029.

[13] E. Stern-Taulats, T. Castán, L. Mañosa, A. Planes, N.D. Mathur, X. Moya, Multicaloric materials and effects, MRS Bull. 43 (2018) 295–299. https://doi.org/10.1557/mrs.2018.72.

[14] J. Liu, N. Scheerbaum, S. Kauffmann-Weiss, O. Gutfleisch, NiMn-Based Alloys and Composites for Magnetically Controlled Dampers and Actuators, Adv. Eng. Mater. 14 (2012) 653–667. https://doi.org/10.1002/adem.201200038.

[15] K. Schlüter, B. Holz, A. Raatz, Principle Design of Actuators Driven by Magnetic Shape Memory Alloys, Adv. Eng. Mater. 14 (2012) 682–686. https://doi.org/10.1002/adem.201200078.

[16] D. Dzekan, A. Waske, K. Nielsch, S. Fähler, Efficient and affordable thermomagnetic materials for harvesting low grade waste heat, APL Mater. 9 (2021) 011105. https://doi.org/10.1063/5.0033970.

[17] A. Waske, D. Dzekan, K. Sellschopp, D. Berger, A. Stork, K. Nielsch, S. Fähler, Energy harvesting near room temperature using a thermomagnetic generator with a pretzel-like magnetic flux topology, Nat. Energy. 4 (2019) 68–74. https://doi.org/10.1038/s41560-018-0306-x.

[18] I. Karaman, B. Basaran, H.E. Karaca, A.I. Karsilayan, Y.I. Chumlyakov, Energy harvesting using martensite variant reorientation mechanism in a NiMnGa magnetic shape memory alloy, Appl. Phys. Lett. 90 (2007) 172505. https://doi.org/10.1063/1.2721143.

[19] T. Gottschall, K.P. Skokov, M. Fries, A. Taubel, I. Radulov, F. Scheibel, D. Benke, S. Riegg, O. Gutfleisch, Making a Cool Choice: The Materials Library of Magnetic Refrigeration, Adv. Energy Mater. 9 (2019) 1901322. https://doi.org/10.1002/aenm.201901322.

[20] T. Gottschall, A. Gràcia-Condal, M. Fries, A. Taubel, L. Pfeuffer, L. Mañosa, A. Planes, K.P. Skokov, O. Gutfleisch, A multicaloric cooling cycle that exploits thermal hysteresis, Nat. Mater. 17 (2018) 929–934. https://doi.org/10.1038/s41563-018-0166-6.

[21] A. Czernuszewicz, J. Kaleta, D. Lewandowski, Multicaloric effect: Toward a breakthrough in cooling technology, Energy Convers. Manag. 178 (2018) 335–342. https://doi.org/10.1016/j.enconman.2018.10.025.

[22] O. Gutfleisch, M.A. Willard, E. Bruck, C.H. Chen, S.G. Sankar, J.P. Liu, Magnetic materials and devices for the 21st century: stronger, lighter, and more energy efficient, Adv Mater. 23 (2011) 821–42. https://doi.org/10.1002/adma.201002180.

[23] T. Krenke, M. Acet, E.F. Wassermann, X. Moya, L. Mañosa, A. Planes, Martensitic transitions and the nature of ferromagnetism in the austenitic and martensitic states of Ni-Mn-Sn alloys, Phys. Rev. B. 72 (2005) 014412. https://link.aps.org/doi/10.1103/PhysRevB.72.014412.

[24] L. Pfeuffer, A. Gràcia-Condal, T. Gottschall, D. Koch, T. Faske, E. Bruder, J. Lemke, A. Taubel, S. Ener, F. Scheibel, K. Durst, K.P. Skokov, L. Mañosa, A. Planes, O. Gutfleisch, Influence of microstructure on the application of Ni-Mn-In Heusler compounds for multicaloric cooling using magnetic field and uniaxial stress, Acta Mater. 217 (2021) 117157. https://doi.org/10.1016/j.actamat.2021.117157.

[25] I.F. Ituarte, F. Nilsén, V.K. Nadimpalli, M. Salmi, J. Lehtonen, S.-P. Hannula, Towards the additive manufacturing of Ni-Mn-Ga complex devices with magnetic field induced strain, Addit. Manuf. 49 (2022) 102485. https://doi.org/10.1016/j.addma.2021.102485.

[26] W. Sun, X. Lu, Z. Wei, Q. Li, Z. Li, Y. Zhang, J. Liu, Multicaloric effect in Ni–Mn–Sn metamagnetic shape memory alloys by laser powder bed fusion, Addit. Manuf. 59 (2022) 103125. https://doi.org/10.1016/j.addma.2022.103125.

[27] A. Milleret, V. Laitinen, K. Ullakko, N. Fenineche, M.M. Attallah, Laser powder bed fusion of (14 M) Ni-Mn-Ga magnetic shape memory alloy lattices, Addit. Manuf. 60 (2022) 103231. https://doi.org/10.1016/j.addma.2022.103231.

[28] M.P. Caputo, A.E. Berkowitz, A. Armstrong, P. Müllner, C.V. Solomon, 4D printing of net shape parts made from Ni-Mn-Ga magnetic shape-memory alloys, Addit. Manuf. 21 (2018) 579–588. https://doi.org/10.1016/j.addma.2018.03.028.

[29] A. Mostafaei, K.A. Kimes, E.L. Stevens, J. Toman, Y.L. Krimer, K. Ullakko, M. Chmielus, Microstructural evolution and magnetic properties of binder jet additive manufactured Ni-Mn-Ga magnetic shape memory alloy foam, Acta Mater. 131 (2017) 482–490. https://doi.org/10.1016/j.actamat.2017.04.010.

[30] F. Nilsén, I.F. Ituarte, M. Salmi, J. Partanen, S.-P. Hannula, Effect of process parameters on non-modulated Ni-Mn-Ga alloy manufactured using powder bed fusion, Addit. Manuf. 28 (2019) 464–474. https://doi.org/10.1016/j.addma.2019.05.029.

[31] C. Lauhoff, A. Fischer, C. Sobrero, A. Liehr, P. Krooß, F. Brenne, J. Richter, M. Kahlert, S. Böhm, T. Niendorf, Additive Manufacturing of Co-Ni-Ga High-Temperature Shape Memory Alloy: Processability and Phase Transformation Behavior, Metall. Mater. Trans. A. 51 (2020) 1056–1061. https://doi.org/10.1007/s11661-019-05608-z.

[32] C. Lauhoff, N. Sommer, M. Vollmer, G. Mienert, P. Krooß, S. Böhm, T. Niendorf, Excellent superelasticity in a Co-Ni-Ga high-temperature shape memory alloy processed by directed energy deposition, Mater. Res. Lett. 8 (2020) 314–320. https://doi.org/10.1080/21663831.2020.1756495.

[33] F. Scheibel, C. Lauhoff, S. Riegg, P. Krooß, E. Bruder, E. Adabifiroozjaei, L. Molina-Luna, S. Böhm, Y.I. Chumlyakov, T. Niendorf, O. Gutfleisch, On the Impact of Additive Manufacturing Processes on the Microstructure and Magnetic Properties of Co–Ni–Ga Shape Memory Heusler Alloys, Adv. Eng. Mater. (2022) 2200069. https://doi.org/10.1002/adem.202200069.

[34] E. Stevens, K. Kimes, V. Chernenko, P. Lázpita, A. Wojcik, W. Maziarz, J. Toman, M. Chmielus, Effect of Homogenization on the Microstructure and Magnetic Properties of Direct Laser-Deposited Magnetocaloric Ni43Co7Mn39Sn11, J. Manuf. Sci. Eng. 142 (2020) 071006. https://doi.org/10.1115/1.4046900.

[35] T. Gottschall, K.P. Skokov, B. Frincu, O. Gutfleisch, Large reversible magnetocaloric effect in Ni-Mn-In-Co, Appl. Phys. Lett. 106 (2015). https://doi.org/10.1063/1.4905371.

[36] R. Kainuma, Y. Imano, W. Ito, Y. Sutou, H. Morito, S. Okamoto, O. Kitakami, K. Oikawa, A. Fujita, T. Kanomata, K. Ishida, Magnetic-field-induced shape recovery by reverse phase transformation, Nature. 439 (2006) 957–960. https://doi.org/10.1038/nature04493.

[37] F. Puglielli, V. Mussi, F. Cugini, N. Sarzi Amadè, M. Solzi, C. Bennati, S. Fabbrici, F. Albertini, Scale-Up of Magnetocaloric NiCoMnIn Heuslers by Powder Metallurgy for Room Temperature Magnetic Refrigeration, Front. Energy Res. 7 (2020) 150. https://doi.org/10.3389/fenrg.2019.00150.

[38] N. Sommer, G. Mienert, M. Vollmer, C. Lauhoff, P. Krooß, T. Niendorf, S. Böhm, Laser Metal Deposition of Fe- and Co-Based Shape-Memory Alloys, Adv. Mater. Res. 1161 (2021) 105–112. https://doi.org/10.4028/www.scientific.net/AMR.1161.105.

[39] V. Petříček, M. Dušek, L. Palatinus, Crystallographic Computing System JANA2006: General features, 229 (2014) 345–352. https://doi.org/10.1515/zkri-2014-1737.

[40] L. Righi, F. Albertini, E. Villa, A. Paoluzi, G. Calestani, V. Chernenko, S. Besseghini, C. Ritter, F. Passaretti, Crystal structure of 7M modulated Ni–Mn–Ga martensitic phase, Acta Mater. 56 (2008) 4529–4535. https://doi.org/10.1016/j.actamat.2008.05.010.

[41] P. Wang, P. Huang, F.L. Ng, W.J. Sin, S. Lu, M.L.S. Nai, Z. Dong, J. Wei, Additively manufactured CoCrFeNiMn high-entropy alloy via pre-alloyed powder, Mater. Des. 168 (2019) 107576. https://doi.org/10.1016/j.matdes.2018.107576.

[42] D.L. Schlagel, R.W. McCallum, T.A. Lograsso, Influence of solidification microstructure on the magnetic properties of Ni–Mn–Sn Heusler alloys, J. Alloys Compd. 463 (2008) 38–46. https://doi.org/10.1016/j.jallcom.2007.09.049.

[43] A. Çakır, L. Righi, F. Albertini, M. Acet, M. Farle, Intermartensitic transitions and phase stability in Ni50Mn50−xSnx Heusler alloys, Acta Mater. 99 (2015) 140–149. https://doi.org/10.1016/j.actamat.2015.07.072.

[44] T. Krenke, E. Duman, M. Acet, E.F. Wassermann, X. Moya, L. Manosa, A. Planes, Inverse magnetocaloric effect in ferromagnetic Ni-Mn-Sn alloys, Nat Mater. 4 (2005) 450–4. https://doi.org/10.1038/nmat1395.

[45] V. Laitinen, A. Sozinov, A. Saren, A. Salminen, K. Ullakko, Laser powder bed fusion of Ni-Mn-Ga magnetic shape memory alloy, Addit. Manuf. 30 (2019) 100891. https://doi.org/10.1016/j.addma.2019.100891.

[46] S. Dhungana, J. Casey, D. Neupane, A.K. Pathak, S. Karna, S.R. Mishra, Effect of Metal-Oxide Phase on the Magnetic and Magnetocaloric Properties of La0.7Ca0.3MnO3-MO (MO=CuO, CoO, and NiO) Composite, Magnetochemistry. 8 (2022) 163. https://doi.org/10.3390/magnetochemistry8120163.

[47] M.C. Brennan, J.S. Keist, T.A. Palmer, Defects in Metal Additive Manufacturing Processes, J. Mater. Eng. Perform. 30 (2021) 4808–4818. https://doi.org/10.1007/s11665-021-05919-6.

[48] A. du Plessis, E. Macdonald, Hot isostatic pressing in metal additive manufacturing: X-ray tomography reveals details of pore closure, Addit. Manuf. 34 (2020) 101191. https://doi.org/10.1016/j.addma.2020.101191.

[49] M. Park, A. Shibata, N. Tsuji, Effect of Grain Size on Mechanical Properties of Dual Phase Steels Composed of Ferrite and Martensite, MRS Adv. 1 (2016) 811–816. https://doi.org/10.1557/adv.2016.230.

[50] P. Entel, M. Siewert, M.E. Gruner, H.C. Herper, D. Comtesse, R. Arróyave, N. Singh, A. Talapatra, V.V. Sokolovskiy, V.D. Buchelnikov, F. Albertini, L. Righi, V.A. Chernenko, Complex magnetic ordering as a driving mechanism of multifunctional properties of Heusler alloys from first principles, Eur. Phys. J. B. 86 (2013) 65. https://doi.org/10.1140/epjb/e2012-30936-9.

[51] E. Wachtel, F. Henninger, B. Predel, Constitution and magnetic properties of Ni-Mn-Sn alloys - solid and liquid state, J Magn Magn Mater. 38 (1983) 305–315. https://doi.org/10.1016/0304-8853(83)90372-4.

[52] V.V. Sokolovskiy, V.D. Buchelnikov, M.A. Zagrebin, P. Entel, S. Sahoo, M. Ogura, First-principles investigation of chemical and structural disorder in magnetic Ni 2 Mn 1 + x Sn 1 − x Heusler alloys, Phys. Rev. B. 86 (2012) 134418. https://doi.org/10.1103/PhysRevB.86.134418.

[53] O. Gutfleisch, T. Gottschall, M. Fries, D. Benke, I. Radulov, K.P. Skokov, H. Wende, M. Gruner, M. Acet, P. Entel, M. Farle, Mastering hysteresis in magnetocaloric materials, Philos. Trans. R. Soc. Math. Phys. Eng. Sci. 374 (2016) 20150308. https://doi.org/10.1098/rsta.2015.0308.

[54] J. López-García, V. Sánchez-Alarcos, V. Recarte, J.A. Rodríguez-Velamazán, I. Unzueta, J.A. García, F. Plazaola, P. La Roca, J.I. Pérez-Landazábal, Effect of high-energy ball-milling on the

magnetostructural properties of a Ni45Co5Mn35Sn15 alloy, J. Alloys Compd. 858 (2021) 158350. https://doi.org/10.1016/j.jallcom.2020.158350.

[55] A.L. Alves, E.C. Passamani, V.P. Nascimento, A.Y. Takeuchi, C. Larica, Influence of grain refinement and induced crystal defects on the magnetic properties of $Ni_{50}Mn_{36}Sn_{14}$ Heusler alloy, J. Phys. Appl. Phys. 43 (2010) 345001. https://doi.org/10.1088/0022-3727/43/34/345001.

[56] S. Singh, P. Kushwaha, F. Scheibel, H.-P. Liermann, S.R. Barman, M. Acet, C. Felser, D. Pandey, Residual stress induced stabilization of martensite phase and its effect on the magnetostructural transition in Mn-rich Ni-Mn-In/Ga magnetic shape-memory alloys, Phys. Rev. B. 92 (2015). https://doi.org/10.1103/PhysRevB.92.020105.

[57] J. Cao, C. Tan, X. Tian, Q. Li, E. Guo, L. Wang, Y. Cao, Effect of Co substitution on magnetic properties of Ni-Mn–Sn magnetic shape memory alloys, Trans. Nonferrous Met. Soc. China. 24 (2014) 1053–1057. https://doi.org/10.1016/S1003-6326(14)63161-7.

[58] D.Y. Cong, S. Roth, L. Schultz, Magnetic properties and structural transformations in Ni–Co–Mn–Sn multifunctional alloys, Acta Mater. 60 (2012) 5335–5351. https://doi.org/10.1016/j.actamat.2012.06.034.

# Supplementary Data

**Additive manufacturing of Ni-Mn-Sn shape memory Heusler alloy – Microstructure and magnetic properties from powder to printed parts**

*Franziska Scheibel, Christian Lauhoff, Philipp Krooß, Stefan Riegg, Niklas Sommer, David Koch, Konrad Opelt, Heiner Gutte, Olena Volkova, Stefan Böhm, Thomas Niendorf, Oliver Gutfleisch*

## 1. Ni-Mn-Sn gas-atomized powder first avalanche, repose angle and cohesion index

The sieved Ni-Mn-Sn powder with a particle size between 20 µm and 50µm shows an average first avalanche angle of 31.8°. For the average ten measurements are performed in a standard drum cell under normal atmosphere. For the measurements a minimum and maximum value of 28.2° and 35.3° are determined.

S 1a) shows the repose angle $\sigma_r$ for different rotation speed 1 to 32 rpm (4 to 132 mms$^{-1}$), and for the different rotation directions: sequence and reverse velocity. The deviation between the two rotation directions is shown in blue. S 1b) shows the corresponding cohesion index [1,2].

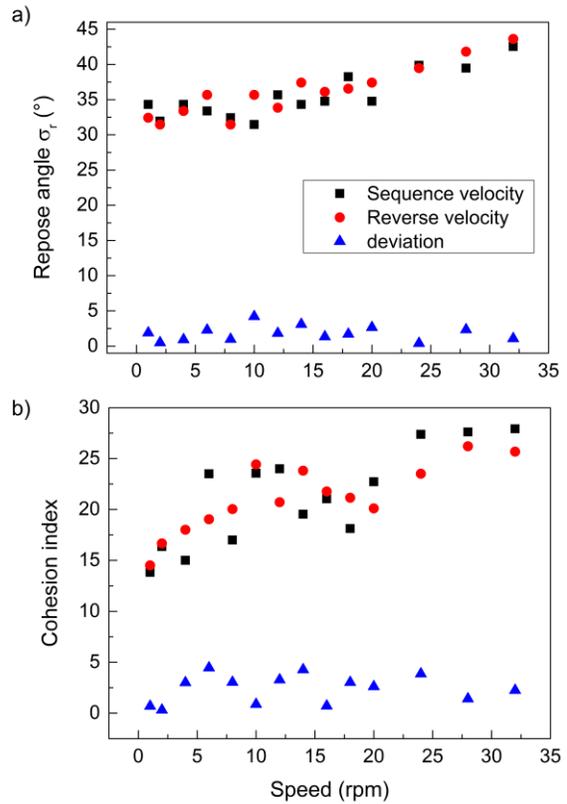

*S 1: a) Repose angle and b) cohesion index of sieved Ni-Mn-Sn powder (20-50 µm) as a function of rotations per minute. Both rotation direction and their deviation are shown.*

# Supplementary Data

## 2. Temperature-dependent x-ray diffraction of as-atomized and heat-treated Ni-Mn-Sn powder

S 2 shows the temperature-dependent x-ray diffraction of the as-atomized (S 2a) and the heat treated (S 2b) Ni-Mn-Sn powder with a particle size between 20 µm and 50 µm. The measurements were carried out during cooling.

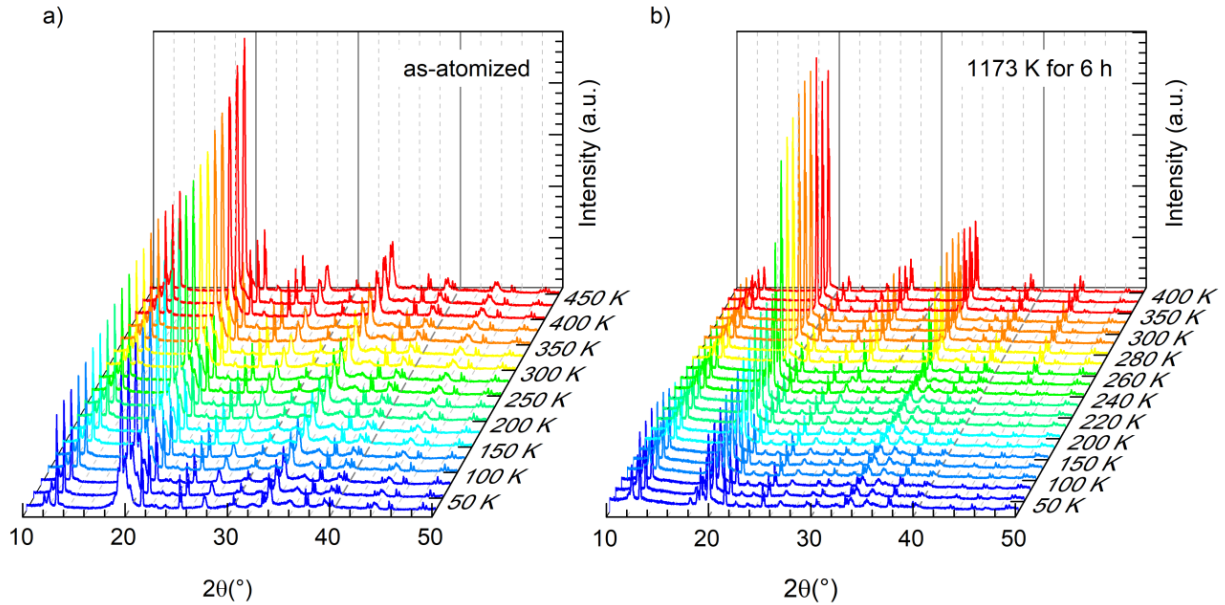

*S 2: Temperature-dependent x-ray diffraction of a) as-atomized and b) heat-treated Ni-Mn-Sn powder with a particle size between 20 µm and 50 µm.*

## 3. Microstructure and composition of heat-treated DED-processed Ni-Mn-Sn

The cross section shown in Fig. 6 was heat-treatment at 1173 K for 24h, the microstructure and chemical composition analysis of the heat-treated cross-section are shown in S 3. The large defects in the area where two tracks are overlapping remain after heat treatment. The magnification of area marked by the red rectangle is shown in S 3c), revealing that both the large pores and the Mn-oxide phase are still present after heat-treatment. The presence of Mn-oxide is confirmed by EDS mapping, shown in S 3e) and f). The magnification of the center of a track, marked by the yellow rectangle, is shown in S 3d). In comparison to Fig. 6d) the same small pores (black contrast) with a size of approximately 1 µm can be still observed. However, the heat-treated DED-processed Ni-Mn-Sn shows no presence of inhomogeneity, the difference in the gray-contrast is related to a grain-contrast. The chemical composition parallel and perpendicular to BD are shown in the line scans in S 3b), indicating a homogeneous composition along both directions. In comparison with the composition in the as-processed state no decrease of the Mn-content parallel to BD can be detected after subsequent heat-treatment. S 3e) and f) show the EDS mapping of selected areas around large pores, the areas are marked in S 3a) by the purple rectangles. The EDS mapping of Mn, O, Sn, and Ni confirm the presence of the Mn-Oxide phase around the large pore.

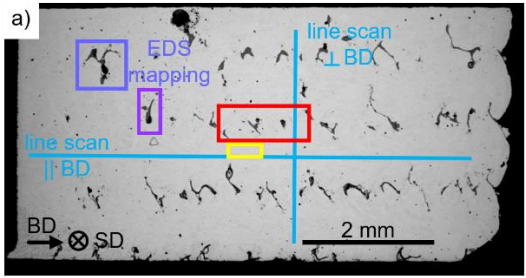

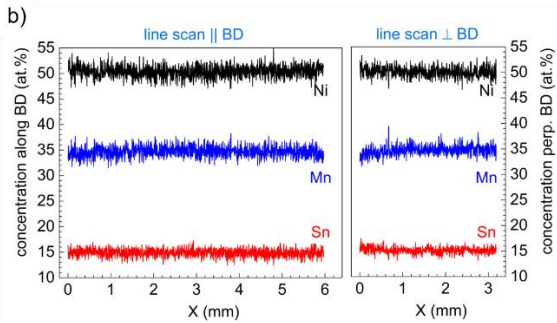

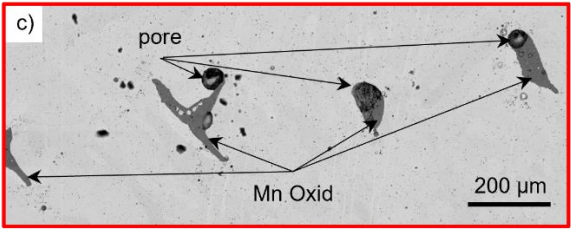

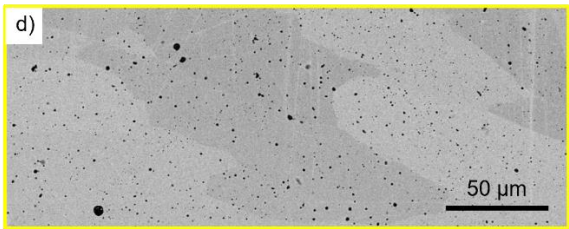

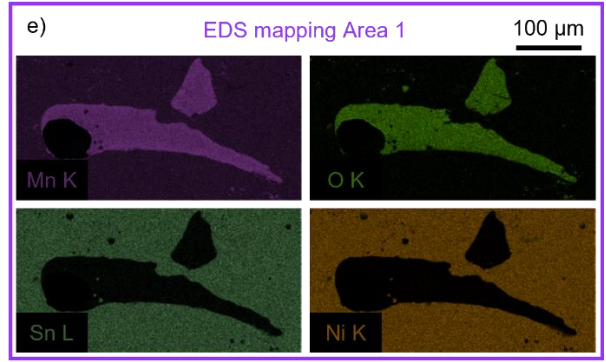

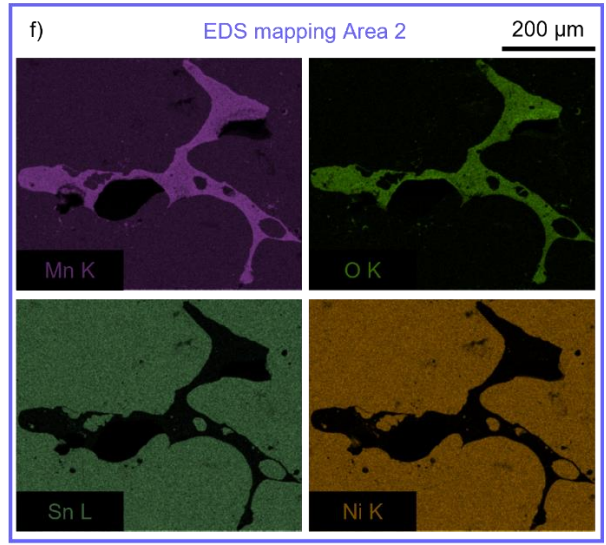

*S 3: a) BSE image and b) EDS line scans of DED-processed and subsequently heat-treated Ni-Mn-Sn. The picture shows the cross-section along BD. c) and d) show the BSE images with higher magnification of the red and yellow marked areas in a), respectively. EDS mapping of Mn, O, Sn, and Ni are shown in e) and f) for selected areas with large pores. The selected areas are marked in a) by purple rectangles.*

# Supplementary Data

**References:**

[1] G. Lumay, F. Boschini, K. Traina, S. Bontempi, J.-C. Remy, R. Cloots, N. Vandewalle, Measuring the flowing properties of powders and grains, Powder Technol. 224 (2012) 19–27.

[2] N. Preud'homme, A. Neveu, F. Francqui, E. Opsomer, N. Vandewalle, G. Lumay, Simulating Powder Bed Based Additive Manufacturing Processes: From DEM Calibration to Experimental Validation, in: 14th WCCM-ECCOMAS Congr., CIMNE, 2021. https://www.scipedia.com/public/Preud'homme_et_al_2021a (accessed October 16, 2022).